\documentclass{emulateapj}

\newcommand{\etal}{\mbox{et~al.\,}}%
\newcommand{\figref}[1]{Figure~\ref{#1}}
\newcommand{\ii}{$i'$}
\newcommand{\zz}{$z'$}
\newcommand{\bb}{$B$}
\newcommand{\vv}{$V$}
\newcommand{\zgal}{$z\!\simeq\!1$}
\newcommand{\sersic}{S\'{e}rsic}

\begin{document}
\title{High Redshift Galaxies in the Hubble Ultra Deep Field}

\author{Nimish   P.  Hathi \\ \emph{Nimish.Hathi@ucr.edu} \\ \emph{Department of Physics \& Astronomy, University of California, Riverside, CA 92521, USA}}

\begin{abstract}

\noindent My dissertation presents results from three recent
investigations in the Hubble Ultra Deep Field (HUDF) focusing on
understanding structural and physical properties of high redshift
galaxies. Here I summarize results from these studies.
\begin{center}
Thesis work conducted at: Arizona State University, Tempe, AZ 85287 

Ph.D. Thesis directed by: Rogier Windhorst and Sangeeta Malhotra 

Ph.D. Degree awarded: August 2008
\end{center}
\end{abstract}


\section{Summary}
In the past decade, space-based and ground-based observations of high
redshift galaxies have begun to outline the process of galaxy
assembly.  The details of that process at high redshifts, however,
remain poorly constrained.  There are two major --- somewhat
contradicting --- scenarios of galaxy assembly and formation.  Many
observational and theoretical predictions favor a hierarchical picture
of galaxy formation, in which galaxies we observe locally were
built-up by a series of mergers from smaller building blocks
\citep[e.g.,][]{ferg04,ryan07}, while in an alternate
`anti'-hierarchical scenario, the most massive galaxies assemble
earlier than their less massive counterparts
\citep[e.g.,][]{heav04,pant07}.  The only way to test these or similar
scenarios, and to constrain their detailed predictions, is to obtain
deep multi-color imaging and study distant galaxies while they undergo
such processes.

This dissertation \citep{hath08c} reports results from three studies
in the Hubble Ultra Deep Field \citep[HUDF;][]{beck06}, the deepest
optical data yet --- which at high redshifts ($z\!\gtrsim\!1$)
corresponds to rest-frame ultraviolet data --- to understand these
distant galaxies.  These three studies focus on understanding
structural and physical properties of galaxies at $z\!\simeq\!4\!-\!6$
and at $z\!\simeq\!1$ using HUDF images and deep grism spectroscopy
from the GRism ACS Program for Extragalactic Science
\citep[GRAPES;][]{pirz04} using Advanced Camera for Surveys (ACS) on
the \emph{Hubble Space Telescope} (\emph{HST}).

\begin{figure}
\epsscale{0.75}
\plotone{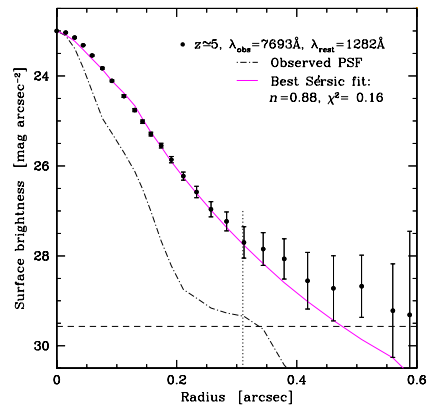}
\caption{Azimuthally averaged radial surface brightness profile
  measured from a composite of 30 $z\!\simeq\!5$ objects
  \citep{hath08a}.  The solid curve is the best-fit S\'{e}rsic
  $r^{1/n}$ profile, with $n$ the S\'{e}rsic index.  The thin
  dot-dashed curve represents the ACS $i'$-band PSF, while the
  horizontal dashed line indicates the surface brightness
  corresponding to the 1\,$\sigma$ uncertainty in the HUDF sky
  background level.  The vertical dotted line marks the radius at
  which the profile starts to deviate significantly from an
  extrapolation of the $r^{1/n}$ profile observed at smaller
  radii.}\label{fig1}
\end{figure}

The HUDF contains a significant number of \bb-, \vv- and \ii-band
dropout objects \citep[e.g.,][]{beck06}, many of which were recently
confirmed to be young star-forming galaxies at $z\!\simeq\!4\!-\!6$.
These galaxies are too faint individually to accurately measure their
radial surface brightness profiles.  Their average surface brightness
profiles are potentially of great interest, since they may contain
clues to the time since the onset of significant galaxy assembly.  In
\citet{hath08a}, we separately co-add \vv-, \ii- and \zz-band HUDF
images of sets of $z\!\simeq\!4,5$ and $6$ objects, pre-selected to
have nearly identical compact sizes and the roundest shapes.  From
these stacked images, we are able to study the averaged radial
structure of these objects at much higher signal-to-noise ratio than
possible for an individual faint object.  In this study, we explore
the reliability and usefulness of a stacking technique of compact
objects at $z\!\simeq\!4\!-\!6$ in the HUDF.  Our results are: (1)
image stacking provides reliable and reproducible average surface
brightness profiles; (2) the shape of the average surface brightness
profile (see, e.g., \figref{fig1}) shows that even the faintest
$z\!\simeq\!4\!-\!6$ objects are \emph{resolved} and that the inner
regions are well represented by disk-like \sersic\ profiles; and (3)
assuming late-type galaxies dominate the population of galaxies at
$z\!\simeq\!4\!-\!6$, as previous \emph{HST} studies have shown for
$z\!\lesssim\!4$, dynamical age estimates for these galaxies from
their profile shapes are comparable with the Spectral Energy
Distribution (SED) ages obtained from the broadband colors.  We also
present accurate measurements of the sky-background in the HUDF and
its associated 1\,$\sigma$ uncertainties.

\begin{figure}
\epsscale{0.75}
\plotone{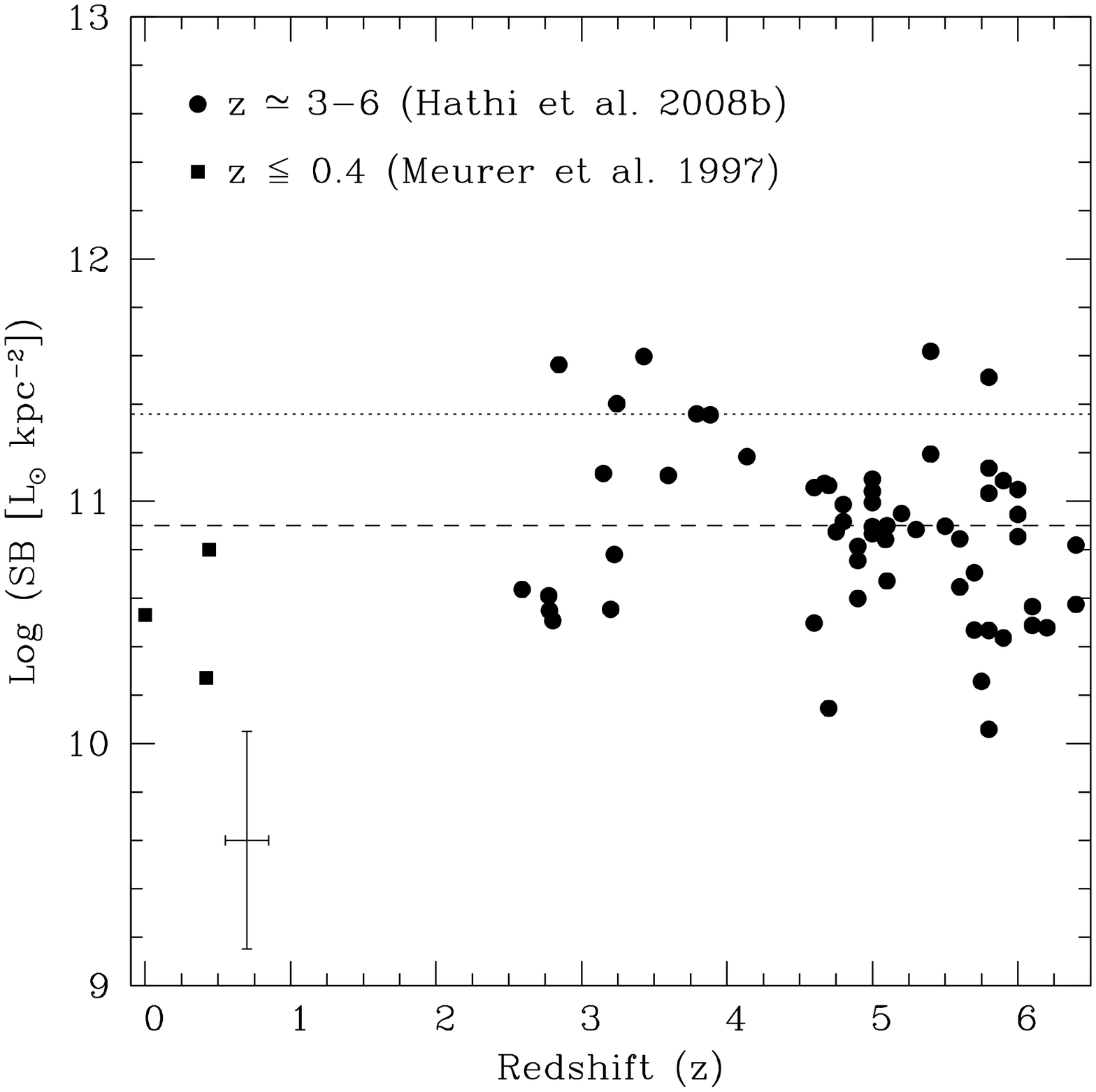}
\caption{Bolometric effective surface brightness (S$_{e}$; L$_{\odot}
  $~kpc$^{-2}$) as a function of redshift \citep{hath08b}.  The filled
  squares at $z\!\simeq\!0-\!0.4$ are measurements of nearby galaxies
  from \citet{meur97}.  The filled circles at $z\!\simeq\!3$ are the
  galaxies from the sample of \citet{meur97} for which we measured
  their surface brightnesses.  The filled circles at
  $z\!\simeq\!4\!-\!6$ are the galaxies in our sample.  The dotted and
  dashed lines correspond to S$_{e,\rm 90}$ (starburst intensity
  limit) and S$_{e,\rm 50}$ (median) of the combine sample,
  respectively.  Uncertainties in the $z\!\simeq\!3\!-\!6$ surface
  brightnesses is shown in the lower left corner.}\label{fig2}
\end{figure}

Starbursts are regions of intense massive star formation that can
dominate a galaxy's integrated spectrum.  By comparing the properties
of starbursts over a wide range of redshifts, we can test whether the
most intense star formation events look the same throughout the
observable history of the universe.  In \citet{hath08b}, we study the
surface brightness properties of 47 spectroscopically confirmed
starburst galaxies at $z\!\simeq\!5\!-\!6$, from the GRAPES project
\citep{malh05,rhoa08}.  We find that the peak star formation intensity
(L$_{\odot}$~kpc$^{-2}$) in starburst galaxies does not vary
significantly from the local universe to redshift $z\!\sim\!6$.  We
arrive at this conclusion through new surface brightness measurements
of 47 starburst galaxies at $z\!\simeq\!5\!-\!6$ (\figref{fig2}),
doubling the redshift range for such observations.  The starburst
intensity limit for galaxies at $z\!\simeq\!5\!-\!6$ agree with those
at $z\!\simeq\!3\!-\!4$ and $z\!\simeq\!0$ \citep{meur97} to within a
factor of a few, after correcting for cosmological surface brightness
dimming and for dust.  The most natural interpretation of this
constancy over cosmic time is that the same physical mechanisms limit
starburst intensity at all redshifts up to $z\!\simeq\!6$ (be they
galactic winds, gravitational instability, or something else).  We
also see two trends with redshift: First, the UV spectral slope
($\beta$) of galaxies at $z\!\simeq\!5\!-\!6$ is bluer than that of
$z\!\simeq\!3$ galaxies, suggesting an increase in dust content over
time.  Second, the galaxy sizes from $z\!\simeq\!3$ to $z\!\simeq\!6$
scale approximately as the Hubble parameter $H^{-1}(z)$.  Thus,
galaxies at $z\!\simeq\!6$ are high redshift starbursts, much like
their local analogs except for slightly bluer colors, smaller physical
sizes, and correspondingly lower overall luminosities.

\begin{figure}
\vspace{0.1in}
\epsscale{0.6}
\plotone{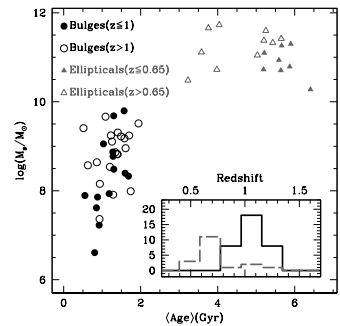}
\caption{Comparison between the ages of late-type bulges at \zgal\ in
  our sample \citep[black circles/lines;][]{hath09} and early-type
  galaxies in GRAPES/HUDF \citep[grey triangles/dashed
    lines;][]{pasq06}. The inset shows the histogram of redshifts for
  both samples with ellipticals peaking around $z\!\simeq\!0.65$.
  Both samples are split with respect to redshift, with solid symbols
  representing the lower redshift subsample.  There is a very
  significant difference between the average age of early-type
  galaxies and late-type galaxy bulges.  Furthermore, the age
  difference is better defined for early-types, suggesting passive
  evolution for these galaxies and a more extended star formation
  history for the late-type bulges.}\label{fig3}
\vspace{-0.2in}
\end{figure}

In \citet{hath09}, we combine HUDF images and GRAPES grism
spectroscopy to explore the stellar populations of 34 bulges belonging
to late-type galaxies at $0.8\!\le\!z\!\le\!1.3$.  The sample is
selected based on the presence of a noticeable 4000~\AA\ break in
their GRAPES spectra, and by visual inspection of the HUDF images.
The HUDF images are used to measure bulge color and \sersic\ index.
The narrow extraction of the GRAPES data around the galaxy center
enables us to study the spectrum of the bulges in these late-type
galaxies, minimizing the contamination from the disk of the galaxy.
We use the low resolution ($R\!\simeq\!50$) SEDs around the
4000~\AA\ break to estimate redshifts and stellar ages.  The SEDs are
compared with models of galactic chemical evolution to determine the
stellar mass, and to characterize the age distribution. We find that,
(1) the average age of late-type bulges in our sample is $\sim$1.3 Gyr
with stellar masses in the range 10$^{6.5}$--10$^{10}$ M$_{\odot}$.
(2) Late-type bulges are younger than early-type galaxies at similar
redshifts and lack a trend of age with respect to redshift, suggesting
a more extended period of star formation (\figref{fig3}).  (3) Bulges
and inner disks in these late-type galaxies show similar stellar
populations, and (4) late-type bulges are better fitted by exponential
surface brightness profiles.  The overall picture emerging from the
GRAPES data is that, in late-type galaxies at \zgal, bulges form
through secular evolution and disks via an inside-out process.



\end{document}